\documentclass[11pt,a4paper]{article}

\usepackage{amsmath}
\usepackage{amsfonts}
\usepackage{amssymb}
\usepackage{graphicx,psfrag,color}
\usepackage{subfigure}
\usepackage{hyperref}

\usepackage{enumerate}

\usepackage[left=2cm,top=2.5cm,right=2cm,bottom=2cm]{geometry}

\begin{document}
\noindent\fbox{\textcolor{blue}{Phys. Lett. B \textbf{737} (2014) 244-247 }}\\

\begin{center}
\Large{\textbf{Observational constraints on variable equation of state parameters of dark matter and dark energy after Planck}} \\[0.5cm]
 
\large{\textcolor{blue}{Suresh Kumar$^*$} \footnote{sukuyd@gmail.com}}, \textcolor{blue}{ Lixin Xu$^{**}$} \footnote{lxxu@dlut.edu.cn}
\\[0.2cm]

\small{
\textit{ $^*$Department of Mathematics, BITS Pilani, Pilani Campus, Rajasthan-333031, India.}}

\small{
\textit{$^{**}$Institute of Theoretical Physics, Dalian University of Technology, Dalian, 116024, P. R. China.}}

\end{center}

\noindent \footnotesize{\textbf{Note: This version of the paper matches the version published in Physics Letters B. The definitive version is available at \href{http://www.sciencedirect.com/science/article/pii/S0370269314006364}{Phys. Lett. B 737 (2014) 244-247}.}}\\

\noindent \small{\textbf{\large{Abstract.}} 
In this paper, we study a cosmological model  in general relativity within the framework of spatially flat Friedmann-Robertson-Walker space-time filled with ordinary matter (baryonic), radiation, dark matter and dark energy, where the latter two components are described by Chevallier-Polarski-Linder equation of state parameters.  We utilize the observational data sets from SNLS3, BAO and {\it Planck}+WMAP9+WiggleZ measurements of matter power spectrum to constrain the model parameters. We find that the current observational data offer tight constraints on the equation of state parameter of dark matter.  We consider the perturbations and study the behavior of dark matter  by observing its effects on CMB and matter power spectra. We find that the current observational data favor the cold dark matter scenario with the cosmological constant type dark energy at the present epoch. \\

\noindent \small{\textbf{\large{PACS.}} 95.35.+d, 95.36.+x, 98.80.Cq



\section{Introduction}
It is not a matter of debate now whether the Universe is accelerating at the present epoch since it is strongly supported by various astronomical probes of complementary nature such as type Ia supernovae data (SN Ia)\cite{1,2}, galaxy redshift surveys \cite{3}, cosmic microwave background radiation (CMBR) data \cite{4,5} and large scale structure \cite{6}. Observations also suggest that there had been a transition of the Universe from the earlier deceleration phase to the recent acceleration phase \cite{7}. We do not have a
fundamental understanding of the root cause of the accelerating expansion of the Universe. We label our ignorance with the term ``Dark Energy" (DE), which is assumed to permeate all of space and increase the rate of expansion of the Universe \cite{8}. On the other hand, the inclusion of DE into the prevailing theory of cosmology has been enormously successful in resolving  numerous puzzles that plagued this field for many years. For example, with prior cosmological models, the Universe appeared to be younger than
its oldest stars.  When DE is included in the model, the problem goes away. The most recent CMB observations indicate that DE accounts for around three fourth  of the total mass energy of the universe
\cite{9,Planck13}. However, the nature of DE is still unknown and various cosmological probes on theoretical and experimental fronts are in progress to resolve this problem. The simplest candidate for the DE is the cosmological constant ($\Lambda$) or vacuum energy since it fits the observational data well. During the cosmological evolution, the cosmological constant has the constant energy density and pressure
with the equation of state (EoS) $w_{de}=p_{de}/\rho_{de}=-1$. However, one has the reason to dislike the cosmological constant since it suffers from the theoretical problems such as the ``fine-tuning" and ``cosmic coincidence" puzzles \cite{10}. Consequently, the dynamic DE models have been studied frequently in the literature. For instance, the Chevallier-Polarski-Linder (CPL) parametrization of the EoS parameter of DE, which  was first introduced in \cite{Chevallier01}, has been frequently constrained with observational data in order to study the nature of dynamic DE (see \cite{Planck13} for recent constraints from Planck).

The $\Lambda$CDM (cosmological constant + cold dark matter) model, which is the standard model in modern cosmology, has been remarkably successful to describe the Universe on large scales. However, it faces persistent challenges from observations on small scales that probe the innermost regions of dark matter halos and the properties of the Milky Way's dwarf galaxy satellites. See \cite{vega11,weinberg13} for reviews on the recent observational and theoretical status of these ``small scale controversies". In this regard, the warm dark matter (WDM)  is a plausible dark matter paradigm, which seems to solve many of small scale discrepancies while being indistinguishable from CDM on larger scales. In particle physics, the keV scale sterile neutrinos are believed to account for WDM.  On the other hand, the fluid perspective of WDM has been investigated in many studies. In \cite{muller05}, the bounds on EoS of DM were investigated using CMB, SN Ia and large scale structure data in the cases of no entropy production and vanishing adiabatic sound speed. In \cite{faber06}, a simple method was suggested for measuring
the EoS parameter of DM that combines kinematic and gravitational lensing data to test the widely adopted assumption of pressureless DM. Following the method, the authors in \cite{serra11} found that the value of the EoS parameter of DM is consistent with pressureless DM within the errors. The authors of \cite{avelino12,cruz13} investigated the ``warmness" of the DM fluid constraining cosmological models with constant EoS parameters of DM and DE by considering non-interacting and interacting scenarios of DM and DE. The authors in \cite{wei13} considered various cosmological models consisting of only DM and DE components by assuming constant and variable EoS parameters of the two components. They  found observational constraints on these models using SN Ia, CMB and BAO data, and concluded that WDM models are not favored over the $\Lambda$CDM model. 

The authors in \cite{Calabrese09} investigated the bounds on EoS parameter of DM using WMAP 5 year data and CMB + SDSS + SNLS data in a cosmological model based on spatially flat Friedmann-Robertson-Walker
space-time filled with ordinary matter (baryonic) and radiation, DE component acting as a cosmological constant and DM component with constant EoS parameter $w_{dm}$. The Friedmann equation in this model reads as
\begin{equation}\label{eq1}
H=H_{0}\sqrt{\Omega_{r}a^{-4}+\Omega_{b}a^{-3}+\Omega_{dm}a^{-3(1+w_{dm})}+
\Omega_{\Lambda}}
\end{equation}
where $a=1/(1+z)$ is scale factor in terms of the redshift $z$; $H_{0}$ is Hubble constant and $\Omega_i=8\pi G \rho_i/(3H_0^2)$ is density parameter for the $i$th component. The authors in \cite{Calabrese09} stressed that in model \eqref{eq1}, the background evolution is completely determined by the EoS of DM, and investigated the properties of DM by studying the behavior on its perturbations. In a recent paper \cite{Lixin13}, the model \eqref{eq1} is constrained with the currently available observational data. In this work, the tighter constraints are obtained on the EoS parameter of DM due to high quality of observational data.

It may be noted that in the studies \cite{Calabrese09,Lixin13}, the DM is characterized by a constant EoS parameter $w_{dm}$ and the DE candidate is the cosmological constant with constant EoS parameter $w_{de}=-1$. However, the choice of constant EoS parameter for DM is too restrictive \cite{wei13}. Similarly, candidature of cosmological constant for DE is not satisfactory as discussed earlier. Therefore, in the present work, we consider the naturally motivated CPL parametrizations for the EoS parameters of DM and DE \cite{wei13}, respectively, given by
\begin{equation}\label{eq2}
w_{de}=w_{dm0}+w_{dma}(1-a),
\end{equation}
\begin{equation}\label{eq3}
w_{de}=w_{de0}+w_{dea}(1-a),
\end{equation}
where $w_{dm0}$, $w_{dma}$, $w_{de0}$ and $w_{dea}$ are constants.

With these CPL forms of EoS of DM and DE, the Friedmann equation can be written as
\begin{equation}\label{eq4}
H=H_{0}\sqrt{\Omega_{r}a^{-4}+\Omega_{b}a^{-3}+\Omega_{dm}f(a)+
\Omega_{de}g(a)}
\end{equation}
where \[f(a)=e^{-3w_{dma}(1-a)}a^{-3(1+w_{dm0}+w_{dma})},\]
\[g(a)=e^{-3w_{dea}(1-a)}a^{-3(1+w_{de0}+w_{dea})}.\]

It is easy to see that the model \eqref{eq1} is retrieved from the model \eqref{eq4} in the particular case $w_{dma}=0$ and $w_{dea}=0$. In the present study, we consider the generalized model \eqref{eq4} and study the constraints on variable EoS parameters of DM and DE by using the currently available observational data from SNLS3, BAO and {\it Planck}+WMAP9+WiggleZ measurements of matter power spectrum. One may observe the strong degeneracy in the background evolution of the DM and DE components. We shall deal with this issue later. Next, we assume that DM component interacts with other components only gravitationally. Since DM is believed to be responsible for the gravitational instability and structure formation in the Universe, we consider the perturbations and study the behavior of DM  by observing its effects on CMB and matter power spectra. Thus, the main objectives of this study include (i) constraining the CPL EoS parameters of DM and DE with the latest observational data (ii) testing the warmness of DM (iii) testing the behavior of DM by observing its effects on CMB and matter power spectra.

\section{Perturbation Equations} \label{sec:PEs}
For a perfect fluid, one has the following perturbation equations for density contrast and velocity divergence in the synchronous gauge
 
 \begin{eqnarray}
 \dot{\delta}_{i}&=&-(1+w_{i})(\theta_{i}+\frac{\dot{h}}{2})+\frac{\dot{w}_{i}}{1+w_{i}}\delta_{i}-3\mathcal{H}(c^2_{s,eff}-c^2_{s,ad})\left[\delta_{i}+3\mathcal{H}(1+w_{i})\frac{\theta_{i}}{k^2}\right],\label{eq:idelta}\\
\dot{\theta}_{i}&=&-\mathcal{H}(1-3c^2_{s,eff})\theta_{i}+\frac{c^2_{s,eff}}{1+w_{i}}k^2\delta_{i}-k^2\sigma_{i}\label{eq:iv},
 \end{eqnarray}
following the notations of \cite{ref:MB}, see also \cite{ref:Hu98,Lixin13}. We have used the following definition of the adiabatic sound speed
\begin{equation}
c^2_{s,ad}=\frac{\dot{p}_{i}}{\dot{\rho}_{i}}=w_{i}-\frac{\dot{w}_{i}}{3\mathcal{H}(1+w_{i})}.
\end{equation}
where $c^2_{s,eff}$ is the effective sound speed in the rest frame of the $i$th fluid. In general, $c^2_{s,eff}$ is a free model parameter, which measures the entropy perturbations through its difference to the adiabatic sound speed via the relation $w_{i}\Gamma_{i}=(c^2_{s,eff}-c^2_{s,ad})\delta^{rest}_{i}$. Thus, $w_{i}\Gamma_{i}$ characterizes the entropy perturbations. Further, $\delta^{rest}_{i}=\delta_{i}+3\mathcal{H}(1+w_{i})\theta_{i}/k^2$ gives a gauge-invariant form for the entropy perturbations. With these definitions, the microscale properties of the energy component are characterized by three quantities, i.e., the EoS parameters $w_{i}$, the effective sound speed $c^2_{s,eff}$ and the shear perturbation $\sigma_i$. In this work, we assume zero shear perturbations for the DM and DE. Since the DM is responsible for the formation of the large scale structure in our Universe, we fix the effective speed of sound $c^2_{s,eff}=0$ for DM in this work. 
\section{Observational constraints} \label{sec:results}
\subsection{Effects of DM parameters on CMB TT and matter power spectra}
Here we analyze that how the DM parameters $w_{dm0}$ and $w_{dma}$ affect the CMB TT and matter power spectra. For this purpose, we fix the other relevant model parameters to their mean values as given in Table \ref{tab:results} and vary one of these two DM parameters $w_{dm0}$ and $w_{dma}$ around its mean value. The effects to the CMB TT power spectrum are shown in Figure \ref{fig:cls}. We see that the positive values of DM parameters $w_{dm0}$ and $w_{dma}$ will decrease the equality time of matter and radiation when the other relevant cosmological model parameters are fixed. Consequently, the amplitudes of the peaks of the CMB are depressed and the positions of the peaks are moved to the right side. On the large scale where $l<10$, the curves are increased when the values of $w_{dm0}$ and $w_{dma}$ are positive due to the integrated Sachs-Wolfe effect. 
\begin{center}
\begin{figure}[htb!]\centering
\includegraphics[width=9.5cm]{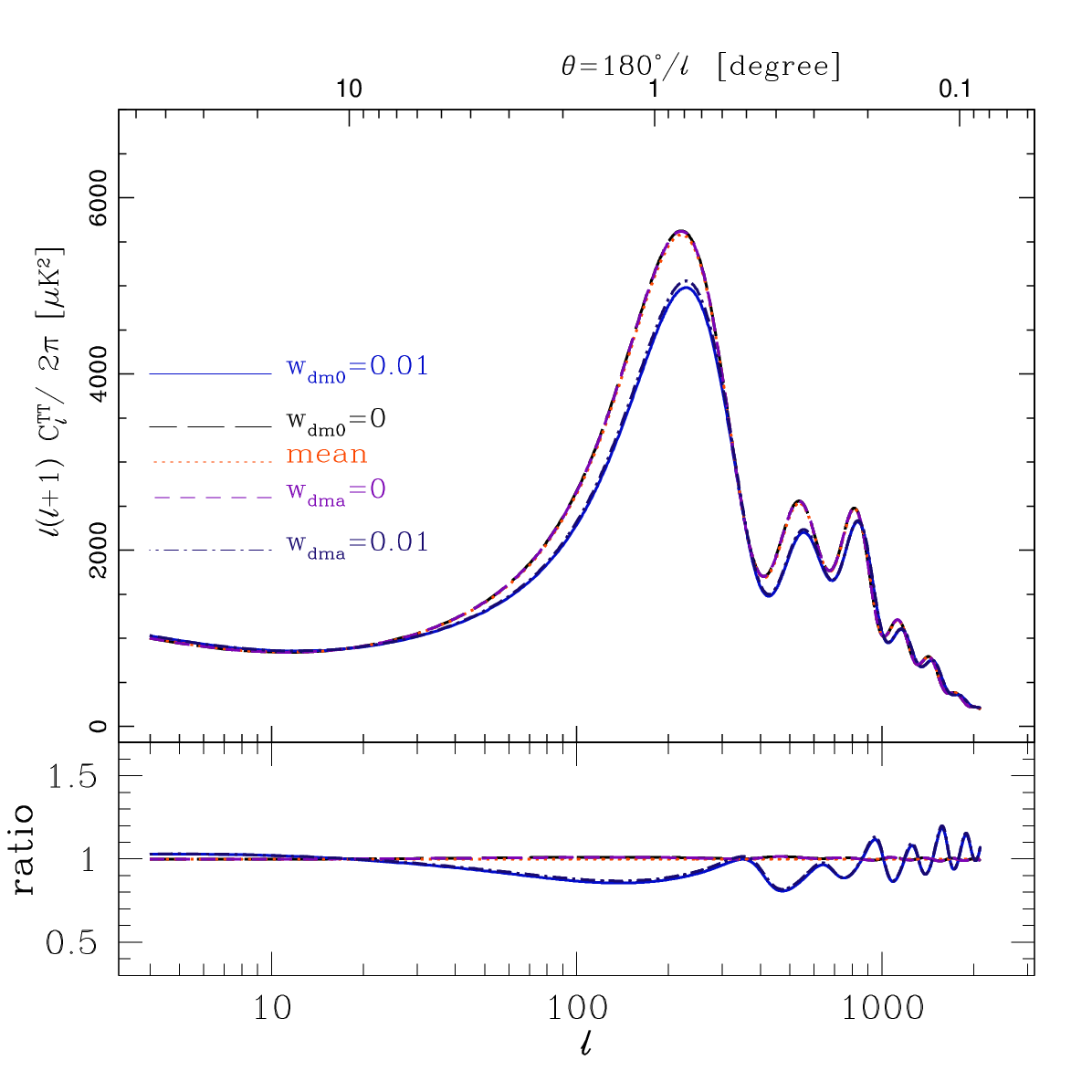}
\caption{The CMB TT power spectrum for different values of DM parameters $w_{dm0}$ and $w_{dma}$, where the other relevant model parameters are fixed to their mean values as shown in Table \ref{tab:results}.}\label{fig:cls}
\end{figure}
\end{center}

The effects of the DM parameters $w_{dm0}$ and $w_{dma}$ on the matter power spectrum are shown in Figure \ref{fig:mp}, where the redshift is fixed to $z=0$. We observe that these effects to the matter power spectrum are similar to the ones as analyzed in Ref. \cite{Lixin13}. The positive values of $w_{dm0}$ and $w_{dma}$ move the matter and radiation equality to earlier times and increase the matter power spectrum.

\begin{center}
\begin{figure}[htb!]\centering
\includegraphics[width=9.5cm]{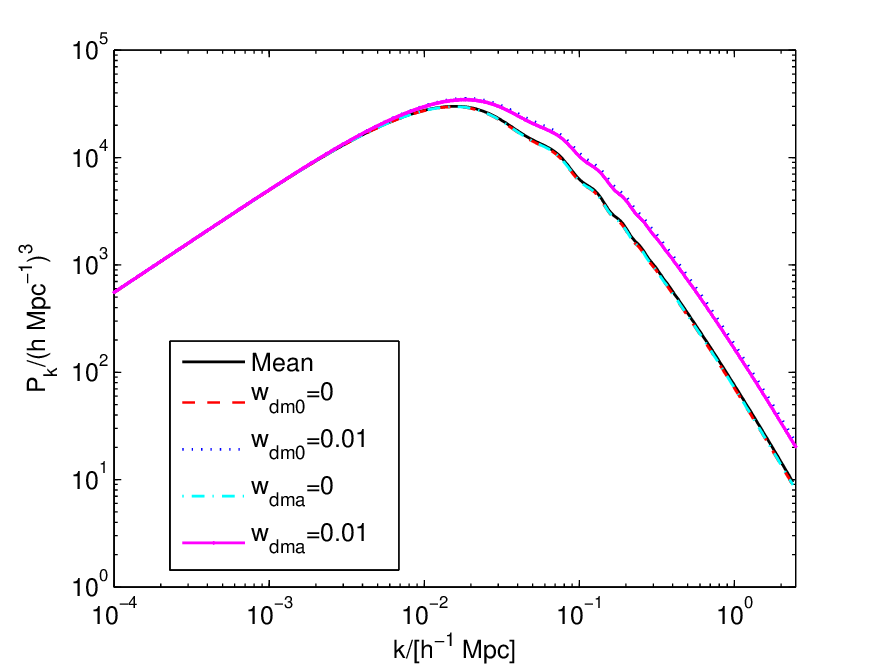}
\caption{The matter power spectrum at redshift $z=0$ for different values of the model parameters $w_{dm0}$ and $w_{dma}$, where the other relevant model parameters are fixed to their mean values as shown in Table \ref{tab:results}.}\label{fig:mp}
\end{figure}
\end{center}

\subsection{Constraints method and fitting results}
We used the observational data points from SNLS3, BAO and {\it Planck}+WMAP9+WiggleZ measurements of matter power spectrum to perform a global fitting to the model parameter space
\begin{equation}
P\equiv\{\omega_{b},\omega_c, \Theta_{S},\tau, w_{dm0}, w_{dma},w_{de0}, w_{dea}, n_{s},\log[10^{10}A_{s}]\}\nonumber
\end{equation}
via the Markov chain Monte Carlo (MCMC) method. We modified the publicly available {\bf cosmoMC} package \cite{ref:MCMC} to include the perturbation evolutions of DM and DE in accordance with the Eqs. (\ref{eq:idelta}) and (\ref{eq:iv}). From the Friedmann equation \eqref{eq4}, one can see that the CPL like DM and DE are degenerated strongly, and through the background evolution information, one can not distinguish the two components. Therefore, in order to obtain tight constraints, one should break this degeneracy by hand, and keep the values of $w_{dm0}$ and $w_{dma}$ in the range of $[0,\infty)$. However, negative values of $w_{dm0}$ are also allowed and discussed in some earlier studies \cite{muller05,Calabrese09,Lixin13}. After assuming the suitable priors on various model parameters (see Table \ref{tab:results}), we ran the code on the {\it Computing Cluster for Cosmos} in eight chains. And it was stopped when the Gelman \& Rubin $R-1$ parameter reached the value $R-1 \sim 0.02$. It guarantees the accurate confidence limits. The results are shown in Table \ref{tab:results}.

In Figure \ref{fig:wigglecontour}, we show one-dimensional marginalized distribution of individual parameters and two-dimensional contours  with $68\%$ C.L. and $95\%$ C.L. for the model parameters under consideration.

\begingroup

\begin{center}
\begin{table}[htb!]\centering
\begin{tabular}{llll}
\hline\hline Parameters &Priors& Mean with errors & Bestfit \\ \hline
$\Omega_b h^2$ &[0.005,0.1]& $0.0219_{-0.00026}^{+0.00026}$ & $0.0219$\\
$\Omega_c h^2$ &[0.01,0.99]& $0.117_{-0.0030}^{+0.0039}$ & $0.120$\\
$100\theta_{MC}$ &[0.5,10]& $1.0413_{-0.00062}^{+0.00062}$ & $1.0411$\\
$\tau$ &[0.01,0.8]& $0.086_{-0.014}^{+0.012}$ & $0.088$\\
$w_{dm0}$ &[0,1]& $0.00067_{-0.00067}^{+0.00011}$ & $0.00029$\\
$w_{dma}$ &[0,1]& $0.00068_{-0.00068}^{+0.000099}$ & $0.00014$\\
$w_{de0}$& $[-5,0]$& $-1.06_{-0.13}^{+0.11}$ & $-1.01$\\
$w_{dea}$&$ [-5,5]$& $0.03_{-0.40}^{+0.68}$ & $-0.40$\\
$n_s$& [0.9,1.1]& $0.959_{-0.0069}^{+0.0069}$ & $0.960$\\
${\rm{ln}}(10^{10} A_s)$ &[2.7,4] & $3.084_{-0.024}^{+0.024}$ & $3.091$\\
\hline
$\Omega_{de}$ &-& $0.720_{-0.011}^{+0.011}$ & $0.717$\\
$\Omega_m$ & -&$0.280_{-0.011}^{+0.011}$ & $0.283$\\
$\sigma_8$ &- &$0.87_{-0.03}^{+0.03}$ & $0.88$\\
$z_{re}$ &-& $10.7_{-1.1}^{+1.1}$ & $11.0$\\
$H_0$ &-& $70.6_{-1.2}^{+1.2}$ & $70.9$\\
${\rm{Age}}/{\rm{Gyr}}$ &-& $13.70_{-0.05}^{+0.05}$ & $13.71$\\
\hline\hline
\end{tabular}\caption{The mean values with $1\sigma$ errors and the best fit values of model parameters, where SNLS3, BAO, {\it Planck}+WMAP9+WiggleZ measurements of matter power spectrum are used.}\label{tab:results}
\end{table}
\end{center}
\endgroup

\begin{center}
\begin{figure}[htb!]\centering
\includegraphics[width=9.5cm]{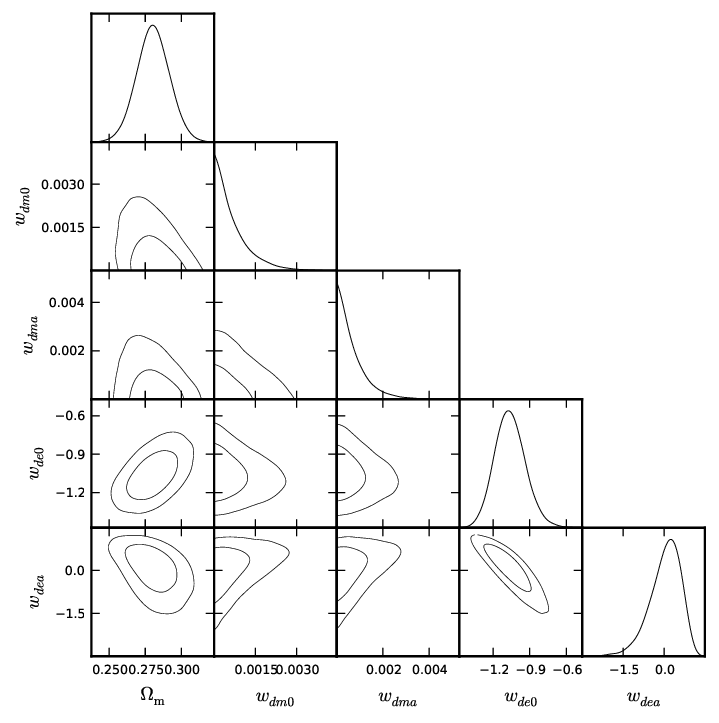}
\caption{The one-dimensional marginalized distribution of individual parameters and two-dimensional contours  with $68\%$ C.L. and $95\%$ C.L.}\label{fig:wigglecontour}
\end{figure}
\end{center}

\section{Summary and conclusions}
From Table \ref{tab:results}, we see that the 1$\sigma$ constraints on $w_{dm0}$ are $0.00067_{-0.00067}^{+0.00011}$, which are tighter in comparison to the constraints obtained in the recent studies \cite{Calabrese09,Lixin13}. Also, the gradual shift of the values of $w_{dm0}$ towards 0 with the advanced and improved observational data shows that the current/future observations are likely to favor the CDM scenario over the WDM. Similarly, the constraints on $w_{de0}$  are  $-1.06_{-0.13}^{+0.11}$. Also, the best fit value of $w_{de0}$ is $-1.01$. Thus, current observational data favor cosmological constant type of DE.  Hence, the $\Lambda$CDM scenario is favored by the currently available observational data  within the framework of the  model considered in this study. 

In the present study, we have investigated the fluid perspective of DM via variable EoS in CPL form by confronting the model with the latest observational data. We have not found any significant deviation from the CDM scenario. The authors of a recent study \cite{Picon}, on the other hand, placed model independent upper limits on the temperature to mass ratio of CDM today using CMB
and matter power spectra observations under the assumptions that DM particle decoupled kinetically while non-relativistic, when galactic scales had not entered the horizon yet, and that their momentum distribution has been Maxwellian since that time. They found that the CDM is quite cold with velocity dispersion smaller than 54 m/s. Thus, the currently available observational data do not provide any compelling evidence against the CDM scenario, and consequently solution to the small scale problems, as discussed in the introduction, remains illusive which could be solved otherwise. We hope that the future observational data would shed light on mysterious nature of DM.\\

\noindent \textbf{Acknowledgements:} S.K. acknowledges the warm hospitality and research facilities provided by the Inter-University Centre for Astronomy and Astrophysics (IUCAA), India where a part of this work was carried out. Further, S.K. acknowledges the financial support from the Department of Science and Technology (DST), India under project No. SR/FTP /PS-102/2011. L. Xu's work is supported in part by NSFC under the Grants No. 11275035 and ``the Fundamental Research Funds for the Central Universities" under the Grants No. DUT13LK01.

\end{document}